\documentclass[oneside,twocolumn,superscriptaddress,eqsecnum,amsmath,showpacs]{revtex4-1}

\usepackage{bm}
\usepackage{amssymb}
\usepackage{amsfonts}
\usepackage{braket}
\usepackage{epsfig}
\usepackage{graphicx}
\usepackage{natbib}
\usepackage{mathdots}

\newcommand{\vep}{\varepsilon}

\newcommand{\nn}{\nonumber \\}

\newcommand{\al}{\alpha}
\newcommand{\be}{\beta}

\newcommand{\de}{\delta}

\newcommand{\ka}{\kappa}
\newcommand{\om}{\omega}

\newcommand{\pd}{\partial}
\newcommand{\ba}{\begin{eqnarray}}
\newcommand{\ea}{\end{eqnarray}}
\renewcommand{\v}[1]{{\bf #1}}
\graphicspath{{figures/}}

\begin{document}

\title{Dynamics of magnon fluid in Dzyaloshinskii-Moriya magnet and \\ its manifestation in magnon-Skyrmion scattering}

\author{Yun-Tak Oh}
\affiliation{Department of Physics, Sungkyunkwan University, Suwon
  440-746, Korea}
\author{Hyunyong Lee}
\affiliation{Department of Physics, Sungkyunkwan University, Suwon 440-746, Korea}
\author{Jin-Hong Park}
\affiliation{Center for Emergent Matter Science, RIKEN, Wako, Saitama 351-0198, Japan}
\author{Jung Hoon Han}
\email[Electronic address:$~~$]{hanjh@skku.edu}
\affiliation{Department of Physics, Sungkyunkwan
University, Suwon 440-746, Korea}
\date{\today}

\begin{abstract}
We construct Holstein-Primakoff Hamiltonian for magnons in arbitrary
slowly varying spin background, for a microscopic spin Hamiltonian consisting of
ferromagnetic spin exchange, Dzyaloshinskii-Moriya
exchange, and the Zeeman term. The Gross-Pitaevskii-type equation for magnon dynamics
contains several background gauge fields pertaining to local
spin chirality, inhomogeneous potential, and anomalous
scattering that violates the boson number conservation. Non-trivial corrections to previous formulas derived in the literature are given. Subsequent mapping to hydrodynamic fields yields the continuity equation and the Euler equation of the magnon fluid dynamics. Magnon wave scattering off a localized Skyrmion is examined numerically based on our Gross-Pitaevskii formulation. Dependence of the effective flux experienced by the impinging magnon on the Skyrmion radius is pointed out, and compared with analysis of the same problem using the Landau-Lifshitz-Gilbert equation.
\end{abstract}
\maketitle

\section{Introduction}
Magnons are quantized waves of spin fluctuations around an ordered
state of spins in a magnet. The theory of their dynamics was
formulated in the 1940s by Holstein and Primakoff\,\cite{holstein1940}
and for this reason they are also known as Holstein-Primakoff\,(HP)
bosons. Magnon dispersion in the ferromagnetic or the
anti-ferromagnetic spin background can be worked out easily from the
HP theory as explained in textbooks on quantum
magnetism\,\cite{auerbach94}.

Modern experimental and theoretical advances call for refinements of
the HP theory.  On the experimental front abundant sightings of the
Skyrmionic spin texture in chiral magnets demand a well-defined
theory of spin waves in the textured spin
background\,\cite{nagaosa2013}. Existing
attempts\,\cite{yehua2011,mochizuki12,yehua13,jiadong2013,iwasaki2014} for theories of
spin dynamics in the Skyrmion background are done in terms of the
Landau-Lifshitz-Gilbert (LLG) equation for the magnetization unit vector $\v n (\v r
, t)$. Despite its obvious strength in direct simulation of the
magnetization dynamics in such complex spin background, LLG approach is not well suited for an intuitive understanding of the low-energy spin dynamics based on the particle picture. Moreover, magnon scattering off the localized Skyrmion is a subject of growing importance and interest as addressed
by several recent works\,\cite{jiadong2013, iwasaki2014}. Here again, a formulation of the problem in terms of the well-established HP boson theory is surprisingly lacking.

As amply demonstrated in several recent theories of
magnon Hall effect\,\cite{katsura2010, murakami2011a, murakami2011b,
murakami2014, lee14}, HP theory conveniently captures magnon
dynamics in a non-trivial spin background in a manner completely analogous to that of
electron dynamics in a non-trivial flux background and band Chern
number. In these approaches, however, the relevant description is within the momentum space picture assuming translationally invariant spin background over some large unit cell. We try in this article to present a thorough, self-contained formulation of magnon dynamics in {\it real space},
suited in particular for adressing the scattering process off a localized object such as a
Skyrmion. The work is thus complementary to both strands of existing theories of spin dynamics
based on LLG equation\,\cite{jiadong2013,iwasaki2014}, and those formulated in momentum space\,\cite{katsura2010, murakami2011a, murakami2011b,murakami2014, lee14}.

In section \ref{sec:magnon-Hamiltonian} we present a general HP Hamiltonian assuming arbitrary smooth spin background of the ferromagnetic exchange Hamiltonian supplemented by Dzyaloshinskii-Moriya (DM) exchange and Zeeman terms. While this formulation was attempted in some earlier works~\cite{tserkovnyak2012,loss2013} and applied in various context of spintronics,  not all of the terms we derive here have been coherently presented thus far. In particular the terms directly following from the DM spin interaction have never been consistently derived to our knowledge. Then in sec. \ref{sec:hydrodynamics} we show how the magnon equation of motion presented in the previous section can be equivalently viewed as a hydrodynamics problem, with a set of hydrodynamic equations pertaining to magnon density and velocity flows. Numerical simulations based on HP theory are in fact equivalent to solving the hydrodynamics version of the magnon problem. We present such numerical result in sec. \ref{sec:numerical}, in the particular case of magnon scattering off a single, localized Skyrmion. Our results partially overlap with an earlier simulation of the same problem~\cite{iwasaki2014}. Here we find a surprising dependence of the magnon scattering angle (and even its sign!) on the Skyrmion radius normalized by the spiral wavelength. This unexpected feature of the magnon-Skyrmion scattering follows from taking thorough consideration of the DM term in the HP formulation. We carry out a numerical check of the prediction using the more conventional LLG approach. It turns out that LLG approach embodies a more complex picture of the scattering dynamics due to the fact that the Skyrmion tends to respond dynamically to the impinging magnon wave in the LLG theory, whereas the HP theory in effect treats it as a static object of infinite mass and internal excitation energies. Despite these differences, we were able to deduce features of the subtle dependence of the magnon scattering on the Skyrmion radius that was unexpected in previous simulations~\cite{jiadong2013,iwasaki2014}. We conclude with a summary and possible future applications of our formulation in sec. \ref{sec:summary}.

\section{Hamiltonian formulation of magnon dynamics}
\label{sec:magnon-Hamiltonian}

The low-energy spin dynamics of a ferromagnet is captured by the
nonlinear sigma model (NL$\sigma$M)

\begin{eqnarray}
  H_0 = \frac{J}{2}\sum_{\mu=1}^d (\pd_\mu \v n ) \cdot (\pd_\mu \v n )
  \label{eq:nlsig_h}
\end{eqnarray}
with the spatial index $\mu$ running over 1 through $d$ in
$d$-dimensional space. When a smooth perturbation is added to the
NL$\sigma$M the new ground state will generally become a slowly varying
spin texture. In light of the recent surge of interests in textured ground states of the
chiral magnet\,\cite{nagaosa2013},
we include the DM interaction,

\begin{eqnarray}
  H_{\rm DM} =D \v n \cdot \bm \nabla \times \v n,
  \label{eq:dm_h}
\end{eqnarray}
as well as the Zeeman term

\ba H_{\rm Z} = - S \v B \cdot \v n \ea
($S=$ size of spin) as the perturbation. Space integration is implicitly assumed in the above expressions.
It is useful for further technical development to point out that the combined spin Hamiltonian $H = H_0 +
H_{\rm DM}+ H_{\rm Z} $ can be organized in the compact form, up to an
irrelevant constant and taking $\kappa=D/J$,

\begin{eqnarray}
  {H \over J} = \frac{1}{2}\sum_{\mu=1}^d \left( \pd_{\mu} \v n -
  \ka \hat{e}_{\mu} \times \v n \right)^2 - {S \over J}  \v B \cdot \v n,
  \label{eq:dm_inserted_h}
\end{eqnarray}
where $\hat{e}_{\mu}$ is the unit vector in each $\mu$-direction. Henceforth
we choose the unit of energy $J=1$ and re-scale the field strength $\v
B$ accordingly. Restriction to $d=2$ is easily achieved by deleting all terms from Eq. (\ref{eq:dm_inserted_h}) pertaining to the spatial direction $z$.

Let us denote the new ground state or some metastable spin
configuration of the Hamiltonian $H$ by $\v n_0$. Given such

\ba \v n_0 = (\sin\theta \cos\phi,\, \sin\theta \sin\phi,\, \cos\theta )\nonumber \ea
one can construct an orthogonal matrix\,$R$ whose elements are

\begin{eqnarray}
  &&R_{\al\be} = 2 m_{\al} m_{\be} - \de_{\al\be},\nn
  &&\v m = \Big( \sin\frac{\theta}{2}\cos\phi,\, \sin\frac{\theta}{2}
  \sin\phi,\, \cos\frac{\theta}{2} \Big).
  \label{eq:rot_mat}
\end{eqnarray}
With this $R$ one can show $R\, \hat{z}  = \v n_0$ and $R\v n_0 =
\hat{z}$. The strategy is to introduce HP bosons in the rotated frame
of reference, as obtained by replacing
$\v n$ in Eq.\,\eqref{eq:dm_inserted_h} by $R \v n$.

After some extensive algebra one finds the Hamiltonian in the new spin basis

\begin{eqnarray}
  H= \frac{1}{2} \sum_{\mu=1}^d
  \left(\pd_{\mu} \v n  - \v a_{\mu} \times \v n
  \right)^2 - S B\, \v n_0 \cdot \v n
  \label{eq:h_rotated}
\end{eqnarray}
where, component-wise, the vector potentials are

\begin{eqnarray}
  && a_{\mu}^1 = - n_x^0(\pd_{\mu} \phi) - \sin\phi(\pd_{\mu}\theta)
  + \ka\, f_{\mu}(\theta,\phi),\nn
  && a_{\mu}^2 = \cos\phi (\pd_{\mu} \theta) - n_y^0(\pd_{\mu}\phi)
  + \ka\, g_{\mu}(\theta,\phi), \nn
  && a_{\mu}^3 = (1-\cos\theta)(\pd_{\mu} \phi) + \ka\,n_{\mu}^0.
  \label{eq:a_with_dm}
\end{eqnarray}
The two functions $f_{\mu}(\theta,\phi)$ and $g_{\mu}(\theta,\phi)$ tied to the DM interaction are

\begin{eqnarray}
  &&f_{\mu}(\theta,\phi)= \Bigg\{\begin{array}{cc}
    -\cos^2\frac{\theta}{2}+\cos 2\phi \sin^2\frac{\theta}{2}&(\mu=x)\\
    \sin 2\phi \sin^2\frac{\theta}{2}&(\mu=y)\\
    n_x^0&(\mu=z)
  \end{array}, \nn
  &&g_{\mu}(\theta,\phi)= \Bigg\{ \begin{array}{cc}
    \sin 2\phi \sin^2\frac{\theta}{2}&(\mu=x)\\
    -\cos^2\frac{\theta}{2} -\cos 2\phi \sin^2\frac{\theta}{2}&(\mu=y)\\
    n_y^0 &(\mu=z)
  \end{array}.
  \label{eq:fg_func}
\end{eqnarray}
All the $\kappa$-dependent terms in the emergent gauge fields $\v a_\mu$ were missed in the literature.

Quadratic magnon Hamiltonian follows from the HP substitution in the rotated spin Hamiltonian \,\eqref{eq:h_rotated}:

\begin{widetext}
\begin{eqnarray}
  {H \over S} &=& \frac{1}{2} \sum_{\mu=1}^d
  \bigl[(\pd_{\mu} \!+\! i a_{\mu}^3) b_{\v r}^{\dag} \bigr]
  \bigl[(\pd_{\mu} \!-\! ia_{\mu}^3) b_{\v r} \bigr] +
  \frac{1}{2} \sum_{\mu=1}^d
  \bigl[(\pd_{\mu} \!-\! i a_{\mu}^3) b_{\v r}\bigr]
  \bigl[(\pd_{\mu} \!+\! ia_{\mu}^3) b_{\v r}^{\dag}\bigr]
  \nn
  &&  + \left( \v B \cdot \v n_0 -\frac{1}{2} \sum_{\mu=1}^d
  [(a_{\mu}^1)^2 \!+\! (a_{\mu}^2)^2]  \right) b_{\v r}^{\dag}b_{\v r}
  - \frac{1}{4} \sum_{\mu=1}^d
  (a_{\mu}^1 \!-\! i a_{\mu}^2)^2 b_{\v r}b_{\v r}
  - \frac{1}{4}\sum_{\mu=1}^d
  (a_{\mu}^1 \!+\! i a_{\mu}^2)^2 b_{\v r}^{\dag} b_{\v r}^{\dag} .
  \label{eq:dm_hp_h}
\end{eqnarray}
\end{widetext}
The size of spin $S$ that serves as an overall constant in the quadratic theory can be set to one. Heisenberg equation of motion for the boson operator is

\begin{widetext}
\ba
   i \hbar \frac{\partial b_{\v r} }{\partial t}
   = -\sum_{\mu=1}^d
    (\pd_{\mu} - i a_{\mu}^3)^2 b_{\v r}
   + \left( \v B \cdot \v n_0 -\frac{1}{2} \sum_{\mu=1}^d
  [(a_{\mu}^1)^2 + (a_{\mu}^2)^2]  \right) b_{\v r}
   - \frac{1}{2} \left( \sum_{\mu=1}^d  (a_{\mu}^1 + i a_{\mu}^2)^2 \right)
   b_{\v r}^{\dag}  .
  \label{eq:eom_boson}
\ea
\end{widetext}
This equation of motion for the magnon operator $b_{\v r}$, combined with Eqs.\,\eqref{eq:a_with_dm} and \eqref{eq:fg_func} for the background gauge fields, gives the governing dynamics of
magnons moving in the arbitrary smooth spin background $\v
n_0$. Relation of the gauge field $\v a^3$ to the background spin $\v n_0$
is

\begin{eqnarray}
(\bm \nabla \times \v a^3 )_\alpha
  &=&{1 \over 2} \vep_{\alpha\mu\nu} \v n_0 \cdot (\pd_{\mu} \v n_0 \times
  \pd_{\nu} \v n_0)
  + \ka (\bm \nabla \times \v n_0 )_\alpha \nn
&& + (1-\cos \theta) (\bm \nabla \times \bm \nabla \phi )_\alpha  .\label{eq:curl-a3}
\end{eqnarray}
The prefactor in the first term on the r.h.s. shows that one Skyrmion
acts as two units of flux quanta for the bosons. Interestingly, the spin vorticity $\bm \nabla \times \v n_0$ also contributes to $\bm \nabla \times \v a^3$ due to the DM interaction. We will see later that this extra contribution has an interesting consequence for the magnon-Skyrmion scattering problem. The final expression on the r.h.s. might be ignored on the ground that $\bm \nabla \times \bm \nabla \phi =0$, but as we will soon see this is not the case with the Skyrmion or the anti-Skyrmion configuration we have in mind.

The anomalous potential, which violates the boson number conservation, has the
connection to the background spin

\ba && - \sum_{\mu=1}^d (a_{\mu}^1 \! + \! i a_{\mu}^2)^2
\nn
&& ~~ =\!  e^{2i\phi} \sum_{\mu=1}^d
  \left[ (\hat{\theta} \!+\! i \hat{\phi}) \cdot ( \pd_{\mu} \v n_0  \!-\! \ka
  \hat{e}_{\mu}\times \v n_0) \right]^2 .
\label{eq:coefficient1}
\ea
Other unit vector fields in the above formula are

\ba \hat{\theta} &=& (\cos\theta \cos\phi, \cos\theta \sin\phi,
-\sin\theta), \nn
\hat{\phi} &=& (- \sin\phi, \cos\phi, 0) , \nonumber \ea
both locally orthogonal to $\v n_0$.

The meaning of the anomalous term becomes clear with the exemplary
spiral spin $\v n_0 = (\cos \kappa z, \sin \kappa z, 0)$ - a
right-handed spiral with the propagation vector $\v k =
(0,0,\kappa)$. We find that Eq. (\ref{eq:coefficient1}) reduces to
$\kappa^2 e^{2i\kappa z}$, and the whole magnon Hamiltonian written in
momentum space becomes

\begin{eqnarray}
  H &=& {1\over 4} \sum_{\v q}
  \begin{pmatrix}
    b_{\v q + \v k}^{\dag} \\ b_{- \v q - 3\v k}
  \end{pmatrix}^T
  {\cal H}_{\v q}
 \begin{pmatrix}
    b_{\v q + \v k} \\ b_{- \v q - 3\v k}^{\dag}
  \end{pmatrix},\nn
{\cal H}_{\v q} &=&
  \begin{pmatrix}
    2\v q^2 + \ka^2 & \ka^2 \\
    \ka^2 & 2\v q^2 + \ka^2 \\
  \end{pmatrix} .
\end{eqnarray}
One recovers the well-known magnon dispersion of the spiral spin,
$\om_{\v q} = |\v q|\sqrt{\v q^2 + \ka^2}$. It is through the anomalous potential that the correct magnon dispersion is recovered, while its omission might have led to an incorrect dispersion $\omega_{\v q} = \v q^2 + \kappa^2 /2$.

Finally, there is an inhomogeneous on-site potential

\ba  \sum_{\mu=1}^d [ (a^1_\mu)^2 + (a^2_\mu)^2 ] =  \sum_{\mu=1}^d (\pd_{\mu} \v n_0  \!-\! \ka \hat{e}_{\mu}\times \v n_0)^2 \ea
arising from local deviation of the spin structure from that of a simple spiral.

The formulation presented in this section is completely general, and applies to arbitrary ground state or metastable spin configurations allowed in the Hamiltonian (\ref{eq:dm_inserted_h}).

\section{Hydrodynamic formulation of magnon dynamics}
\label{sec:hydrodynamics}

Magnon dynamics is typically viewed as the solution of
Eq. (\ref{eq:eom_boson}) with some characteristic frequency corresponding to the energy of magnon excitation. In an inhomogeneous environment of the spin $\v n_0$, however, it will be useful to have a complementary picture in real space, quite like that provided by the hydrodynamic formulation of superfluid dynamics~\cite{khalatnikov}, which follows from writing down the corresponding magnon Lagrangian and substituting

\ba b = \sqrt{\rho} e^{i\eta}. \ea
Variation of the Lagrangian with respect to the density
$\rho$ and the phase $\eta$ yields hydrodynamic equations. Another
way, paralleling the derivation of Gross-Pitaevskii equation in
superfluids\,\cite{khalatnikov}, is to replace the boson operator $b$
by its average $\langle b\rangle = \sqrt{\rho}e^{i\eta}$ in the
Heisenberg equation of motion (\ref{eq:eom_boson}). Same hydrodynamic relations are obtained from both approaches.

The boson number is no longer conserved in the continuity equation due to the anomalous potential,

\ba
{ \partial \over \partial t } \rho + \bm \nabla \cdot \left(\rho \v v\right) = \rho F_i , \label{eq:continuity eq}
\ea
where $F_i$ follows from

\ba - e^{-2 i \eta} \left( \sum_{\mu=1}^d ( a_\mu^1 \!+\! i a_\mu^2)^2  \right) \equiv F_r + i F_i . \label{eq:FiFrgauge}  \ea
The other piece of hydrodynamics is provided by the Euler equation of the flow vector $\v v = 2(\bm \nabla \eta - \v a^3 )$ given by

\ba
D_t \v v= \v v \times \v {\cal  B} - \bm \nabla p.
\ea
The material derivative $D_t\equiv \partial_t +\v v \cdot \bm \nabla $ is familiar from hydrodynamics. The emergent magnetic field ${\cal B}$ responsible for the Lorentz force on the magnon is defined by

\ba {\cal B} \equiv -{1\over 2} \bm \nabla \times \v v =\bm \nabla \times \v a^3 -\bm \nabla \times \bm \nabla \eta . \label{eq:emergent-B}\ea
Readers can refer to Eq. (\ref{eq:curl-a3}) for the definition of $\bm \nabla \times \v a^3$.
We will not immediately identify ${\cal B}$ with the curl $\bm \nabla \times \v a^3$ because in some situations the phase singularity in the boson field can lead to $\bm \nabla \times \bm \nabla \eta \neq 0$. Allowing for the time-dependent background spin (which we do not do in this paper) will induce the emergent electric field on the r.h.s. of the Euler equation as well. The quantum pressure $p$ is given by

\ba
p= - {\bm \nabla^2 \sqrt{\rho} \over \sqrt{\rho} }   \!+\! 2  B\, n_0^z \!+\!F_r \!-\! \sum_{\mu=1}^d \left[ (a_\mu^1)^2 \!+\!(a_\mu^2)^2 \right]  .
\ea

There is a tendency in the existing literature~\cite{tserkovnyak2012, loss2013, iwasaki2014,kovalev14,garst14} to
ignore effects arising from $a^1_\mu$ and $a^2_\mu$ in viewing the real-space magnon dynamics, presumably based on the observation that they appear in the magnon Hamiltonian with one
more spatial derivative than $a^3_\mu$. On the other hand, we
already saw that their omission leads to an erroneous result even in the well-known test case. Instead the relevance of the anomalous term has to be addressed carefully, for each specific background spin configuration.

In the next section we take up in particular the problem of magnon scattering off a single
localized Skyrmion based on our HP formulation. This problem is of great topical interest and has already received some attention in the literature~\cite{jiadong2013,iwasaki2014}. The direct numerical approach based on LLG equation in these papers however makes it difficult to interpret the scattering process in terms of magnon particles. Solving the scattering problem analytically together with the anomalous term is
impossible\,\onlinecite{iwasaki2014}. Our method of choice is the numerical
integration of the magnon dynamics, as given in Eq.\,(\ref{eq:eom_boson}).

\section{Simulation of magnon scattering}
\label{sec:numerical}

\subsection{Simulation based on HP theory}
For numerical simulation based on Eq. (\ref{eq:eom_boson}) we work in two dimensions,
external field $\v B = B \hat{z}$, $B>0$, and the DM interaction
constant $\kappa >0$ that favors a right-handed spiral. Under these
parameter values the stable Skyrmion configuration is the one whose
spins point up far from the core, execute a right-handed spiral turn
upon approach to the origin, where it points downward, $-\hat{z}$. A
model spin configuration fulfilling all these requirements is ($r^2 = x^2 + y^2$)

\begin{eqnarray}
  \v n_0 = \Bigg( -\frac{2Ry}{r^2 + R^2},~~ \frac{2Rx}{r^2 + R^2},~~
  \frac{r^2 - R^2}{r^2 + R^2} \Bigg).
  \label{eq:sk_config}
\end{eqnarray}
The Skyrmion number of this configuration is $N=-1$,
qualifying it as an anti-Skyrmion. The HP Hamiltonian
in the presence of a single anti-Skyrmion is

\begin{eqnarray}
H  &=& \sum_{\mu=1}^2 [ (\pd_{\mu} +i a_{\mu}^3) b^\dag_{\v r} ][
(\pd_{\mu}- i a_{\mu}^3) b_{\v r} ] - G(r) b_{\v r}^{\dag}b_{\v r} \nn
&& + {1\over 2} e^{-2 i \phi} F(r) b_{\v r}b_{\v r}+{1\over 2} e^{2 i \phi} F(r) b_{\v r}^{\dag}b_{\v r}^{\dag}, \label{eq:dm_mag_hp_h}
\end{eqnarray}
where

\ba
F(r)&=& \frac{2\ka R (\ka R-2)r^2}{(R^2 + r^2)^2}, \nn
G(r) &=& {R^2(\kappa R-2)^2 + \kappa^2 r^4 \over (R^2+r^2)^2} + B \frac{R^2-r^2}{R^2 + r^2} .
  \label{eq:coeff_sk}
\end{eqnarray}
The $\v a^3$ and its curl is tricky as it contains a singular contribution,

\ba
\v a^3 &=& \left(\frac{-2yR^2 }{r^2 (r^2 + R^2)},
    \frac{2xR^2 }{r^2 (r^2 + R^2)} \right)  \nn
&& ~~~ +  \kappa \left(\frac{-2R y}{r^2 + R^2},
    \frac{2R x}{r^2 + R^2} \right),\nn
\left(\bm \nabla \times \v a^3\right)_z &=& { 4 R^2( \kappa R - 1 ) \over ( r^2 +R^2)^2} + 4\pi \delta (\v r) .
\ea
The delta-function flux can be removed through the singular phase transformation of the boson field $b_{\v r} \rightarrow b_{\v r} e^{2i \phi}$ where $\phi$ is the azimuthal angle in the plane. The new boson field obeys the Hamiltonian

\begin{eqnarray}
H  &=& \sum_{\mu=1}^2 [ (\pd_{\mu} +i a_{\mu}^3) b^\dag_{\v r} ][
(\pd_{\mu}- i a_{\mu}^3) b_{\v r} ] - G(r) b_{\v r}^{\dag}b_{\v r} \nn
&& + {1\over 2} e^{2 i \phi} F(r) b_{\v r}b_{\v r}+{1\over 2} e^{-2 i \phi} F(r) b_{\v r}^{\dag}b_{\v r}^{\dag}, \label{eq:dm_mag_hp_h}
\end{eqnarray}
where $F$ and $G$ remain the same as before, but $\v a^3$ and its curl becomes

\ba
\v a^3 &=&
\left(-\frac{2y(\ka R -1)}{r^2 + R^2},\, \frac{2x(\ka R -1)}{r^2 + R^2}
\right)
 ,\nn
\left(\bm \nabla \times \v a^3\right)_z &=& { 4 R^2( \kappa R - 1 ) \over ( r^2 +R^2)^2} .\label{eq:gauge transf a_3 in skyrm}
\ea
In the numerical treatment we will work in this singularity-free gauge.

It is customary to regard the two-dimensional integral of $(\bm \nabla
\times \v a^3)_z$ divided by $2\pi$ as the effective flux experienced
by the magnon quasiparticle. In such picture each Skyrmion
serves as

\ba \Phi = 2(\kappa R-1) \nonumber \ea
units of flux quanta, spread over the distance of the radius $R$. In contrast to electron motion which sees
the Skyrmion as a quantized flux object carrying one flux
quantum~\cite{nagaosa2013}, magnons see it carrying a variable flux
that can even change sign with the Skyrmion radius!

\begin{figure}[htbp]
\includegraphics[width=0.5\textwidth]{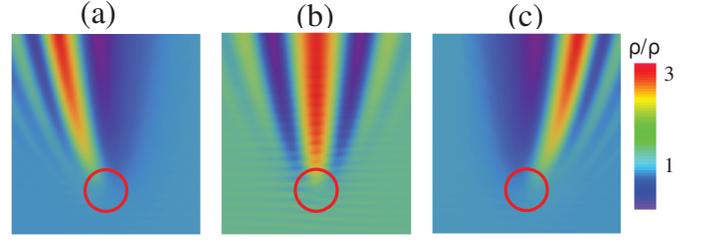}
\caption{(color online) Snapshots of scattered magnon waves for $\tilde{\kappa} = \kappa R$ equal to (a) 0.5, (b) 1.0, and (c) 1.5. Shown are color plots of the magnon density $\rho=|b|^2$ normalized by the incoming magnon density $\rho_0$. Skyrmion is indicated by a red circle of radius $R$. The magnon trajectory changes direction at $\tilde{\kappa}=1$ as anticipated from the total flux consideration in the HP theory.}
\label{fig:HP-simulation}
\end{figure}

Let us rescale all the fields into dimensionless form convenient for numerical analysis,

\begin{eqnarray}
F(\tilde{r}) &=& \frac{2 \tilde{\kappa} (\tilde{\kappa} -2)\tilde{r}^2 }{
    (1+\tilde{r}^2)^2},\nn
G(\tilde{r}) &=& {(\tilde{\kappa} -2)^2 +\tilde{\kappa}^2 \tilde{r}^4 \over (1+\tilde{r}^2)^2}+{\tilde B}\,\frac{1-\tilde{r}^2}{1+\tilde{r}^2}, \nn
\tilde{ \v a}^3 &=& \left(- \frac{2\tilde{y}(\tilde{\ka} -1)}{1+\tilde{r}^2},
\, \frac{2\tilde{x}(\tilde{\ka} -1)}{1+\tilde{r}^2}
\right),
  \label{eq:eg_sk}
\end{eqnarray}
where $\tilde{x}\!=\!x/R$, $\tilde{y}\!=\!y/R$, $\tilde{r} \!=\! r/R$, and
$\tilde{\ka} \!= \!\ka R$ and $\tilde{B} \!=\! B R^2$ are the two dimensionless
parameters. The equation of motion in dimensionless form becomes

\begin{eqnarray}\label{eq:reduced-EOM}
  i\frac{\partial b_{\tilde{\v r}} }{\partial \tilde{t}}
  = - \Bigg[
   (\tilde{\pd}_{\mu} - i \tilde{a}_{\mu}^3)^2
  + G(\tilde{r}) \Bigg] b_{\tilde{\v r}}
 + e^{-2i\phi} F(\tilde{r}) b_{\tilde{\v r}}^*,\nn
 \label{eq:reduced-GP}
\end{eqnarray}
where the rescaled time $\tilde{t} = t/\hbar$ is also dimensionless (Recall that we already set $J=1$ earlier). The tilde can be removed now without confusion.

Hydrodynamic relations following from Eq. (\ref{eq:reduced-GP}) are

\ba
\partial_t \rho + \bm \nabla \cdot (\rho \v v ) = - 2\rho F \sin [ 2(\eta\!+\! \phi ) ] , \label{eq:continuity-equation}
\ea
and

\ba
&& ~~~~~~~ ~~~~~~ D_t \, \v v = \v v \times {\cal B} - \bm \nabla p , \nn
&& p = 2 \left(- {\bm \nabla^2 \sqrt{\rho} \over \sqrt{\rho}} - G+ F\cos
\left[ 2 (\eta \! + \! \phi)\right]\right) .
\ea

We solved Eq. (\ref{eq:reduced-EOM}) using the Runge-Kutta method. Forced spin wave is generated at the bottom of the rectangular simulation grid with the space-time dependence

\ba
b(x,y<y_0 ,t)=b_0 e^{ i k y - i \omega_k t } .
\ea
The frequency $\omega_k \!=\! k^2\! +\! B$ matches the dispersion in
free space far from the Skyrmion, and $y_0$ is a suitable cutoff much less than the total length of the grid. We then closely monitor the scattering processes as the magnons collide with the (static) Skyrmion.

By far the most interesting findings of our investigation was the variation of the scattering angle as $\tilde{\kappa}=\kappa R$ crosses the threshold value 1, whereupon the scattering direction changes sign as shown in Fig. \ref{fig:HP-simulation}. This is a stark prediction of our HP theory, in departure from previous considerations based either on LLG theory~\cite{jiadong2013,iwasaki2014} or on the magnon theory without taking into careful account of the DM term~\cite{tserkovnyak2012,loss2013}. In the following subsection we return to the conventional LLG calculation to verify to what extent, if at all, the predictions of the HP theory is born out.

\subsection{Simulation based on LLG theory}
Lattice version of the LLG equation,

\ba
\dot{\v n}_{i} &=& \v n_i \times  \v B^{\rm eff}_i - \alpha \v n_i \times \left(\v n_i \times  \v B^{\rm eff}_i \right) , \nn
\v B^{\rm eff}_i &=& \sum_{j \in i} \left(\v n_{j}+ a \kappa \v n_{j}\times \hat{\v e}_{j} \right) + a^2 \v B,
\ea
is solved on a large grid of lattice spacing $a$, with the initial spin configuration given by Eq. (\ref{eq:sk_config}). It was found useful to first run the simulation with a large Gilbert constant $\alpha$ of order one to relax the Skyrmion to its true equilibrium configuration, which differs in detail from the model form (\ref{eq:sk_config}), at a given $B$ and $\kappa$. Once an optimal Skyrmion configuration is reached in this way, we turn on the spin wave at the bottom of the boundary to start the scattering process.

Several different Skyrmion radii $R$ were considered in the simulation while simultaneously maintaining the same ratio to the magnon wavelength, $kR$. A detailed dependence of the spin wave scattering angle on this dimensionless quantity $kR$ was reported earlier~\cite{iwasaki2014}, whereas our focus here is with another another dimensionless number, $\kappa R$. Variation of this number can be achieved in practice with the magnetic field $B$, which would choose a different optimal Skyrmion radius $R$.

\begin{figure}[htbp]
\includegraphics[width=0.5\textwidth]{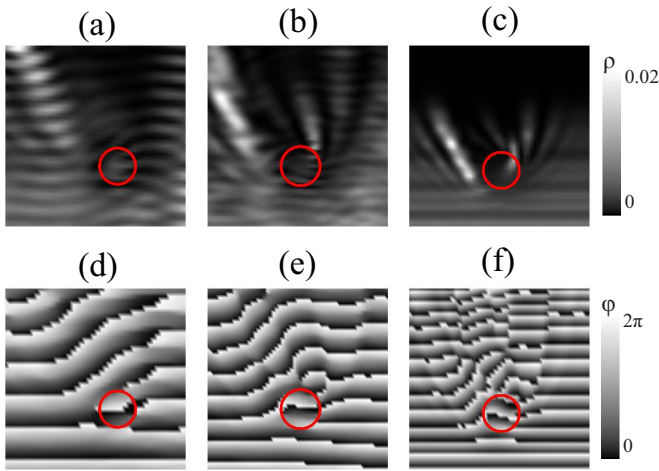}
\caption{(color online) Snapshots of scattered spin waves in the LLG calculation for $\kappa R$ equal to (a,d) 0.9, (b,e) 1.5, and (c,f) 2.0. In the top row are plots of the spin wave density $\rho = \sqrt{( \delta n_x )^2 + (\delta n_y )^2 }$. See the main text for the definition of $\delta \v n = (\delta n_x , \delta n_y)$. Bottom row shows the phase $\phi = \arctan (\delta n_y /\delta n_x )$.  Skyrmion positions and their sizes are indicated by red circles. A clear sign of skew scattering at $\kappa R=0.9$ is completely gone at $\kappa R= 2.0$.}
\label{fig:LLG-simulation}
\end{figure}

Figure \ref{fig:LLG-simulation} shows patterns of scattered spin waves for varying Skyrmion radii $R$. The background spin texture has been removed by taking the time average $\langle \v n_i \rangle $ over a single period of the incoming spin wave, which is then subtracted from the snapshot at a particular instant $\v n_i$ to obtain $\delta \v n_i = \v n_i - \langle \v n_i \rangle$. The resulting profile has only planar components $\delta \v n_i = (\delta n_x , \delta n_y)$ whose magnitude and phase are plotted in Fig. \ref{fig:LLG-simulation}. A small amount of displacement of the Skyrmion~\cite{iwasaki2014} during the cycle does not affect the average $\langle \v n_i \rangle$ significantly.

At the smaller radius $\kappa R=0.9$, overall scattering behavior are analogous to that of magnons for $\kappa R < 1$.  At $\kappa R=1.5$, we see a second scattering channel beginning to open up shown as a white, vertical jet in Fig. 2(b). At $\kappa R= 2.0$, this second channel is pointing at an angle opposite to the first one (Fig. 2(c)), while the phase profile (Fig. 2(f)) becomes nearly isotropic with respect to the incoming spin wave direction. An attempt to carry out LLG calculation at even larger radius $R$, in the hope of finding the reversed scattering direction, failed because stabilizing such a large Skyrmion requires $B$ for which Skyrmion is no longer the most stable phase (instead spiral spin state wins out). To further complicate the analysis, low-energy excitation modes are easily created for larger Skyrmions which subsequently interact with the incoming spin wave.

In short, our LLG simulation does give a hint that spin wave scattering angle has a significant dependence on the reduced Skyrmion radius $\kappa R$, in accordance with anticipations of the HP theory. To observe the proposed effect one can tune the magnetic field over a range in which isolated Skyrmions are the most favored states~\cite{nagaosa2013}. Field-dependent variations of the magnon Hall effect and possible reversal of its sign will be a telltale sign of our effect.

\section{Summary and Discussion}
\label{sec:summary}

Much of this paper dealt with the formal theory of magnon dynamics from Hamiltonian and hydrodynamics perspectives. Despite the long history of the theory of magnon dynamics dating back to 1940s, its real-space formulation with regard to the underlying spin Hamiltonian containing the Dzyaloshinskii-Moriya interaction has never been rigorously derived in the form presented in Sec. \ref{sec:magnon-Hamiltonian}.

The spin Hamiltonian (\ref{eq:dm_inserted_h}) used in this paper is known to host interesting topological phases of spins called the Skyrmion lattice in two dimensions\,\cite{han2010, yu10} and the hedgehog lattice in three\,\cite{jh11, vish06}. The dynamical equation derived in this paper can be used for large-scale simulation of magnon dynamics in these topological phases. The hydrodynamic formulation developed in Sec. \ref{sec:hydrodynamics} can subsequently provide intuitive interpretation of magnon dynamics simulation as movements of magnon fluid.

In this paper we studied one such problem - magnon scattering off a localized Skyrmion - numerically in detail. Magnons do see a localized Skyrmion as a source of flux quanta responsible for skew scattering, in accordance with several earlier observations~\cite{jiadong2013,loss2013,iwasaki2014}, but the net flux seen by the magnon deviates from the two units of flux quanta anticipated in earlier theories due to an important correction from the DM interaction, as in Eq. (\ref{eq:curl-a3}). Our analysis suggests that the net flux seen by the incident magnon may even change its sign at the critical Skyrmion radius $\kappa R_c = 1$. Indeed magnon dynamics simulation based on our HP Hamiltonian did revealed such effect. Unfortunately, the LLG simulation presented a more complex picture due to various low-energy excitations of the Skyrmion induced by the very magnons which scatter off of it. Nevertheless we were able to show that a Skyrmion of increasingly larger radius has less pronounced skew scattering of magnons in rough accordance with the prediction of the HP theory. We anticipate that the magnon Hall effect due to Skyrmions will have a more complex dependence on the external magnetic field (which controls the Skyrmion radius to some extent) than the topological Hall effect experienced by the electrons.

\acknowledgments
This work is supported by the NRF grant\,(No. 2013R1A2A1A01006430). YTO was supported by Global PhD Fellowship Program through the National Research Foundation of
Korea (NRF) funded by the Ministry of Education (2014H1A2A1018320). JHH acknowledges useful discussion with professor Naoto Nagaosa on related topics, and wishes to thank professor Patrick Lee and other members of the condensed matter theory group at MIT for hospitality during his sabbatical stay.

\bibliographystyle{apsrev}

\bibliography{reference}

\end{document}